\documentclass[aps, twocolumn, pre]{revtex4-2}
\usepackage{graphicx, natbib, titlesec, paralist, amsmath,amssymb, color, hyperref}

\def\*#1{\mathbf{#1}}
\begin{document}

\title{Statistical mechanics of the flexural Ising model in $D$ dimensions}
\author{Abigail Plummer}
\affiliation{Department of Mechanical Engineering, Boston University, Boston, MA 02215, USA}
\date{\today}

\begin{abstract}
A generalization of the compressible Ising model in which spins are hosted on an elastic $D$-dimensional lattice embedded in $d>D$ dimensions is studied. Two critical systems interact when temperature is tuned to the Ising transition point, as a freely-fluctuating thermalized crystalline membrane in its flat phase is already critical. Noting that the upper critical dimension of both the membrane and Ising model is $D_{\mathrm{uc}}=4$, renormalization group recursion relations are found by expanding in $\epsilon=4-D$. The coupling between spin and elastic degrees of freedom is shown to be a relevant operator for the physical case of a two-dimensional membrane fluctuating in three-dimensional space ($D=2$, $d=3$), which suggests that the thermalized membrane and Ising systems become more strongly coupled at long wavelengths. The coupling is irrelevant when the difference between the space dimension and membrane dimension is greater than twelve. 
\end{abstract}

\maketitle

\section{INTRODUCTION}

In the 1970s, the impact of elastic lattice deformations on the critical behavior of the Ising model was studied in detail using the then new tools of renormalization group \cite{sak1974critical, wegner1974magnetic, bergman1976critical, de1976coupling, bruno1980renormalization}. Researchers sought to understand whether it was reasonable to use Ising spins on a rigid lattice to understand real materials, which have finite elastic constants. Renormalization group calculations suggest that compressibility \textit{can} affect phase behavior, in some cases changing the second order transition into a first-order transition or altering the critical exponents. The appearance and magnitude of these changes depends strongly on the dimension/symmetry of the spin lattice and the choice of ensemble. 

In the late 1980s, these same renormalization group tools were applied to investigate $D$-dimensional crystalline membranes fluctuating in $d$ dimensions, with ${d>D}$ \cite{aronovitz, david1988crumpling, paczuski1988landau}. Consistent with previous calculations and simulations \cite{nelson1987fluctuations, kantor1987crumpling}, a flat phase with a diverging bending rigidity was shown to exist over a range of temperatures with scale-dependent elastic moduli. Further inspired by the experimental realization of free-standing sheets of graphene in 2004 \cite{geim2007rise}, renormalization group studies of thermalized membranes have continued to yield insights into the mechanical behavior of fluctuating atomically-thin materials \cite{katsnelson2007graphene, kovsmrlj2016response, le2018anomalous, shankar2021thermalized}

In recent years, inspired by nanoscale and macroscale metamaterials as well as buckled atomically thin monolayers \cite{seffen2006mechanical, oppenheimer-PRE-92-052401-2015,
faber-advancedScience-7-2001955-2020, liu2022snap, shohat2022memory, paulsen2024mechanical, seixas2016multiferroic, molle2017buckled, hanakata2017two}, a model consisting of a periodic array of buckled bistable nodes embedded in a 2D crystalline membrane with periodic boundaries has been studied with simulations and theory \cite{plummer2020buckling, hanakata2022anomalous, plummer2022curvature}. Bistable nodes are formed by locally dilating the lattice---as the amount of dilation increases, it becomes energetically favorable for a node to buckle up or down out of the plane, trading stretching energy for bending energy. The buckled nodes couple to one another via the elastic energy of the host membrane such that neighboring buckled ``spins" prefer to be antialigned at zero temperature, similar to an Ising antiferromagnet. 

To establish a more precise connection with Ising antiferromagnets, molecular dynamics simulations of the buckled dilation array model were conducted at finite temperature by \citet{hanakata2022anomalous}. Upon increasing the temperature, thermal fluctuations disrupt the antiferromagnetic ordering of the buckled nodes. This behavior can be characterized as a phase transition with the orientation of the buckled nodes defining a staggered magnetization order parameter, and critical exponents can be measured using finite size scaling. Many of the exponents measured in simulation are consistent with the Ising universality class, including susceptibility exponent $\gamma$, magnetization exponent $\beta$, and correlation length exponent $\nu$. However, the value of the specific heat critical exponent $\alpha$ is higher than expected ($\alpha/\nu=0.068 \pm 0.018$, whereas the 2D Ising model predicts $\alpha/\nu=0$), hinting that the elasticity of the host lattice could be influencing critical behavior. 

Most strikingly, the buckled dilation array has a diverging coefficient of thermal expansion that changes sign, from negative to positive, at the Ising critical point. Models of pristine membranes (as well as real materials, such as graphene \cite{yoon2011negative}) display a negative coefficient of thermal expansion---as temperature increases, their projected area shrinks due to entropic effects. This behavior is observed in the discrete buckled dilation model when far from the Ising critical temperature. Close to the critical temperature, the buckled nodes experience fluctuations in their ordering at all scales, and the disordered packing of nodes produces a swelling of the membrane that dominates the entropic shrinkage and leads to a positive coefficient of thermal expansion. This tunability of the coefficient of thermal expansion is unusual and desirable in engineered materials \cite{boatti2017origami, lopez2017tailoring}.

\begin{figure}[htp]
\begin{center}
\includegraphics[scale=0.25]{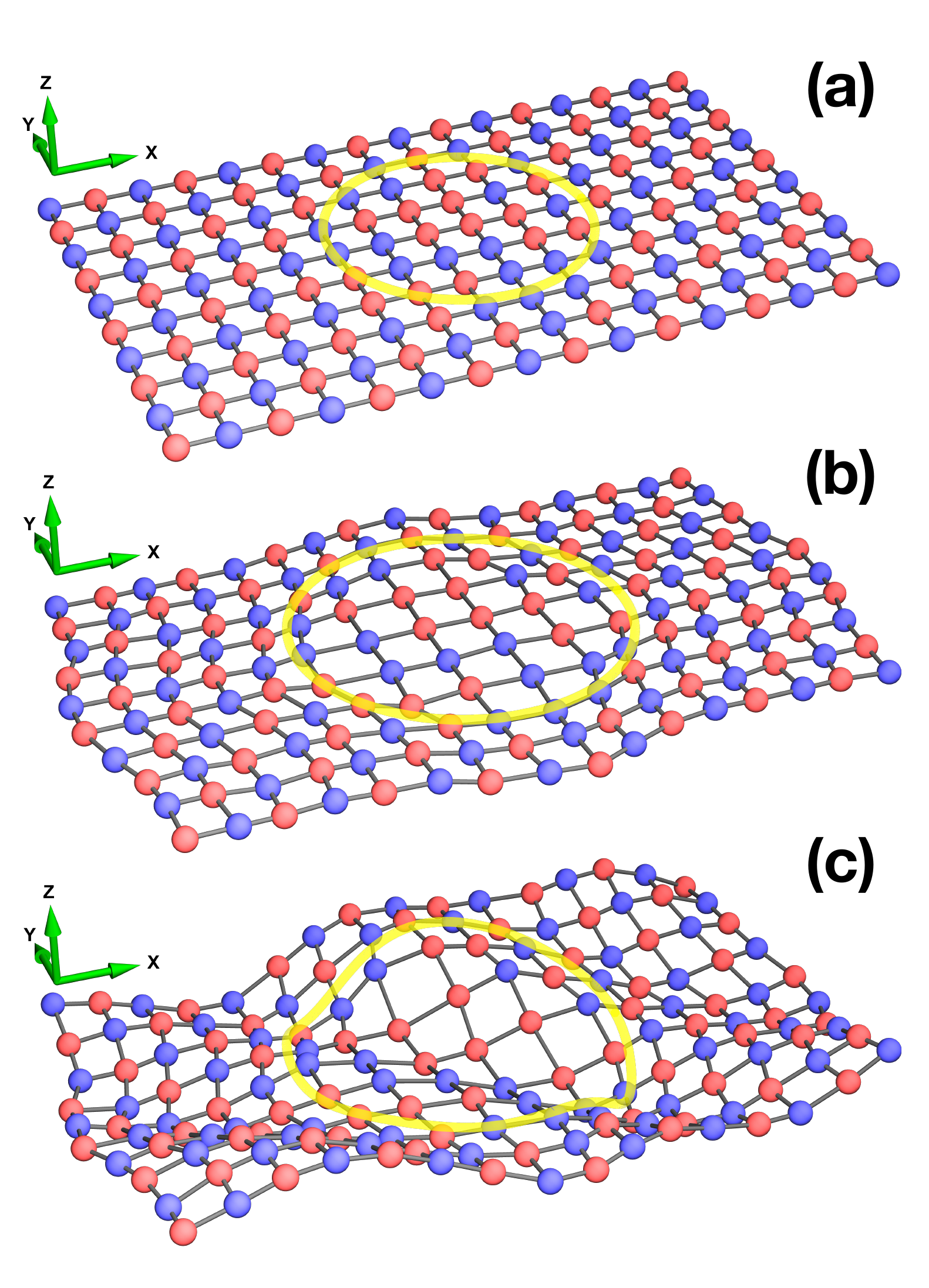}
\end{center}
\caption{{\label{fig:schematic} Schematic showing the effects of introducing compressibility into a 2D Ising antiferromagnet embedded in three dimensions ($D=2, d=3$). (a) A rigid Ising antiferromagnet with uninterrupted antiferromagnetic order outside of the yellow circle and disorder inside the yellow circle. Interacting neighbors are connecting by lines. (b) An Ising antiferromagnet with in-plane compressibility. The disordered region inside the circle expands due to a coupling between the magnetization and the strain. (c) A flexural Ising antiferromagnet with compressibility. The spin lattice expands in the disordered region and can fluctuate in three dimensions. 
}}
\end{figure}

In order to understand the anomalous thermal expansion of buckled dilation arrays, a novel generalization of the compressible Ising model called the ``flexural Ising model" was introduced \cite{hanakata2022anomalous}. The flexural Ising model couples an Ising order parameter to a thin elastic sheet with both in-plane and out-of-plane fluctuations. The schematic in Fig. \ref{fig:schematic} clarifies the differences between the rigid Ising model (Fig. \ref{fig:schematic}a), the compressible Ising model (Fig. \ref{fig:schematic}b), and the flexural Ising model (Fig. \ref{fig:schematic}c). The flexural Ising model successfully describes the thermal expansion observed in dilation array models. However, its implications for critical phenomena remain unexplored. Understanding the fixed points of this model under renormalization group flows could help explain the apparent deviation of the specific heat exponent $\alpha$ from the Ising universality class value observed in simulations. Moreover, it provides an exciting opportunity to study the interaction between two simultaneously critical systems with different origins: a thermalized membrane with critical fluctuations over a wide range of temperatures and an Ising system that can be tuned to its critical point. 

Here, renormalization group methods are applied to the flexural Ising model. This report investigates the relevance of the coupling between flexural and magnetic degrees of freedom and searches for new fixed points that are not present in standard compressible Ising models. The analysis finds that the coupling is relevant for two-dimensional sheets embedded in three-dimensional space but does not identify a new fixed point towards which the Hamiltonian is expected to flow.




\section{EFFECTIVE HAMILTONIAN OF THE FLEXURAL ISING MODEL}
\label{sec:heff}
The effective Hamiltonian for an Ising model hosted on an isotropic $D$-dimensional lattice allowed to fluctuate in $d$ dimensions has three contributions: terms from the free energy of a (rigid) Ising model, terms from the elastic free energy of a fluctuating crystalline membrane, and a term coupling the two systems \cite{hanakata2022anomalous}.
\begin{equation}
\mathcal{H}_{\mathrm{eff}}=\mathcal{H}_{\mathrm{Ising}}+\mathcal{H}_{\mathrm{elastic}}+\mathcal{H}_{\mathrm{couple}}.
\label{eq:totalen}
\end{equation}
The Ising contributions are given by the Landau-Ginzberg Hamiltonian,
\begin{equation}
\mathcal{H}_{\mathrm{Ising}}=\int d^D x \left(\frac{K}{2} (\nabla m)^2+\frac{r}{2} m^2 +u m^4 \right),
\end{equation}
where $m$ is a scalar order parameter. For an Ising ferromagnet, $m$ is the magnetization of the system. For an Ising antiferromagnet (as pictured in Fig. \ref{fig:schematic}), $m$ is staggered magnetization, which tracks the prevalence of checkerboard order.

The elastic contributions are
\begin{equation}
\mathcal{H}_{\mathrm{elastic}}=\int d^D x \left(\frac{\kappa}{2} |\nabla^2 \*f|^2 + \mu u_{ij}^2 + \frac{\lambda}{2} u_{kk}^2\right). 
\end{equation}
Here, $u_{ij}$ is the strain tensor, defined in terms of $u_i$, a $D$-dimensional vector of in-plane displacements and $\*f$, a ${d-D\equiv d_c}$-dimensional vector of out-of-plane displacements, such that
\begin{equation}
u_{ij}= \frac{1}{2} \left( \frac{\partial u_j}{\partial x_i} + \frac{\partial u_i}{\partial x_j} + \frac{\partial \*f}{\partial x_i} \cdot \frac{\partial \*f}{\partial x_j} \right).
\end{equation}
Note that higher order terms quadratic in in-plane displacements, $\frac{\partial u_k}{\partial x_i}\frac{\partial u_k}{\partial x_j}$, are absent \cite{nelson2004statistical}.

The magnetic and elastic degrees of freedom are coupled by the lowest order term allowed by symmetry \cite{larkin1969phase, sak1974critical}.
\begin{equation}
\mathcal{H}_{\mathrm{couple}}=\int d^D x g m^2 u_{kk}.
\end{equation}
As depicted in Fig. \ref{fig:schematic}, buckled dilation arrays have $g>0$, as the antiferromagnetic state has a smaller projected area than the ferromagnetic state at $T=0$, consistent with the positive peak observed in the coefficient of thermal expansion \cite{hanakata2021thermal}. See, e.g., refs. \cite{wagner1970elasticity, baker1970effects, catterall1992scaling, shokef2011order, ruiz2015ripples, garcia2023buckling, mukherjee2022statistical} for related models considering couplings between Ising-like order parameters and surfaces of various kinds.

Since the energy functional is linear and quadratic in the in-plane displacement fields $u_i$, standard techniques are used to isolate these terms and trace over them \cite{larkin1969phase, sak1974critical, nelson1987fluctuations, le2018anomalous}. Details are given in Appendix A. This procedure results in an energy functional that depends only on $m$ and $\*f$. The effective Hamiltonian of Eq. \ref{eq:totalen} can then be written in real space as
{\small
\begin{align}
\mathcal{H}&= \frac{v}{L^D} \left(\int d^D x m^2 \right)^2+ \int d^D x \left( \frac{K}{2}\left(\nabla m\right)^{2}+\frac{r}{2} m^{2}+ \tilde{u} m^{4} \right) \nonumber\\ &+ \int^\prime d^D x  \bigg[ \frac{\kappa}{2}|\nabla^2 \*f|^2+ \frac{\mu}{4}\left(P_{il}^{T} \partial_{i} \*f \cdot \partial_{j} \*f\right)\left(P_{jk}^{T} \partial_{k} \*f \cdot \partial_{l} \*f\right)\nonumber \\&+\frac{\mu}{4}\left(1- \frac{2b}{DB}\right)\left(P_{ij}^{T} \partial_{i} \*f \cdot \partial_{j} \*f\right)^{2}+  w m^2 \left(P_{ij}^T \partial_i \*f\cdot \partial_j \*f \right)\bigg] , 
\label{eq:energy}
\end{align}
}where primed integrals exclude the zero mode of the strain tensor, $P_{ij}^T$ is the transverse projection operator, $B= (2\mu+D \lambda)/D$ is the $D$-dimensional bulk modulus, and 
\begin{align*}
v&=\frac{g^2}{2}\left( \frac{1}{2 \mu+\lambda}-\frac{D}{ 2\mu+D\lambda}\right) \\&\hspace{2 em}=\frac{g^2}{2 B}\left(\left(1+\frac{2(D-1)\mu}{D B} \right)^{-1} -1\right), \\
\tilde{u}&=u- \frac{g^2}{2(2\mu+\lambda)},\\
b&=\frac{\mu (2 \mu+ D \lambda)}{2\mu+\lambda}=\frac{\mu D B}{(2\mu+\lambda)},\\
w&=\frac{g \mu}{2 \mu+\lambda}. 
\end{align*}

The interesting, highly nonlocal coupling proportional to $v$ emerges from the integration of the uniform component of the strain tensor. This term is also present for compressible Ising models without flexural phonons and can affect critical behavior \cite{sak1974critical, de1976coupling, bergman1976critical, bruno1980renormalization}. Note that since the bulk and shear modulus must be positive, $v$ must be negative. Similar nonlocal couplings also appear in constrained Ising models \cite{rudnick1974renormalization} and clamped fluctuating surfaces \cite{hanakata2021thermal, shankar2021thermalized}. 

The coupling proportional to $w$, on the other hand, is unique to the flexural Ising model and allows flexural phonons and the local magnetization to interact. Understanding the impact of this term on critical phenomena is the primary objective of this work. 

\section{BEHAVIOR AWAY FROM THE CRITICAL POINT}\label{sec:rescl}
Prior to studying behavior close to the Ising critical point using renormalization group methods, the flexural Ising model is examined above and below $T_c$. Arguments suggest that magnetic degrees of freedom do not affect the critical behavior of the flexural phonons in these limits and vice versa. 


At temperatures far below the Ising critical temperature, the magnetization $m$ is approximately $\bar{m}$, a nonzero constant. The energy simplifies to
\begin{align}
\mathcal{H}&\approx \frac{r}{2} L^D \bar{m}^{2}+ (\tilde{u}+v) L^D \bar{m}^{4}+ \nonumber \\
&+\int^\prime d^D x \bigg[\frac{\kappa}{2}|\nabla^2 \*f|^2 +\frac{\mu}{4}\left(P_{il}^{T} \partial_{i} \*f \cdot \partial_{j} \*f\right)\left(P_{jk}^{T} \partial_{k} \*f \cdot \partial_{l} \*f\right)\nonumber \\&\hspace{5 em}+\frac{\mu}{4}\left(1- \frac{2b}{DB}\right)\left(P_{ij}^{T} \partial_{i} \*f \cdot \partial_{j} \*f\right)^{2} \bigg],
\end{align}
where $L^D$ is the $D$-dimensional area of the undeformed membrane. No terms couple magnetization to out-of-plane fluctuations---the term proportional to $w$ is zero due to the restriction that the $\*q=0$ mode is excluded from the integral. Thus, quantities such as $\langle \*f(\*q)\cdot \*f(-\*q) \rangle$ are expected to take typical values for a pristine thermalized membrane \cite{nelson2004statistical}. 

The in-plane fluctuations of the host lattice can affect the magnetic degrees of freedom through the definitions of $\tilde{u}$ and $v$, as described in past studies of the compressible Ising model \cite{bruno1980renormalization}. If $\tilde{u}+v <0$ (possible for large coupling $|g|$ and/or small bulk modulus $B$), higher order terms in $m$ are required for stability. Including new terms could allow for a first order transition: If $u_6 m^6$ with $u_6>0$ is added, for example, Landau theory predicts an abrupt jump in $\bar{m}$ at $r=\frac{(\tilde{u}+v)^2}{2 u_6}$. 

At temperatures far above the critical point, $m$ is small and quartic terms can be neglected. Gradients in $m$ are also neglected, assuming that fluctuations are smaller than the coarse-graining scale. 
\begin{align}
\mathcal{H}& \approx \int^\prime d^D x \bigg[\frac{\kappa}{2}|\nabla^2 \*f|^2+\frac{\mu}{4}\left(P_{il}^{T} \partial_{i} \*f \cdot \partial_{j} \*f\right)\left(P_{jk}^{T} \partial_{k} \*f \cdot \partial_{l} \*f\right)\nonumber 
\\&\hspace{5 em}+\frac{\mu}{4}\left(1- \frac{2b}{DB}\right)\left(P_{ij}^{T} \partial_{i} \*f \cdot \partial_{j} \*f\right)^{2} \nonumber \\&\hspace{5 em}+ \left(w P_{ij}^T \partial_i \*f\cdot \partial_j \*f + \frac{r}{2} \right)m^2 \bigg], 
\end{align}
The magnetization is traced out at every point in space and constants are neglected.
\begin{align}
\mathcal{H}& \approx \int^\prime d^D x \bigg[\frac{\kappa}{2}|\nabla^2 \*f|^2\nonumber +\frac{\mu}{4}\left(P_{il}^{T} \partial_{i} \*f \cdot \partial_{j} \*f\right)\left(P_{jk}^{T} \partial_{k} \*f \cdot \partial_{l} \*f\right)\nonumber 
\\&\hspace{5 em}+\frac{\mu}{4}\left(1- \frac{2b}{DB}\right)\left(P_{ij}^{T} \partial_{i} \*f \cdot \partial_{j} \*f\right)^{2}\nonumber \\&\hspace{5 em}+ \frac{k_B T }{2 a_0^D} \ln \left( 1+ \frac{2 w}{r} P_{ij}^{T} \partial_{i} \*f \cdot \partial_{j} \*f \right) \bigg], 
\end{align}
where $a_0$ is the lattice constant. Upon expanding the logarithm to quadratic order, assuming $\frac{2 w}{r} P_{ij}^{T} \partial_{i} \*f \cdot \partial_{j} \*f \ll 1$, the energy becomes
\begin{align}
\mathcal{H}& \approx \int^\prime d^D x \bigg[\frac{\kappa}{2}|\nabla^2 \*f|^2 +\frac{\mu}{4}\left(P_{il}^{T} \partial_{i} \*f \cdot \partial_{j} \*f\right)\left(P_{jk}^{T} \partial_{k} \*f \cdot \partial_{l} \*f\right)\nonumber \\&\hspace{5 em}+\frac{\mu}{4}\left(1- \frac{2b}{DB}\right)\left(P_{ij}^{T} \partial_{i} \*f \cdot \partial_{j} \*f\right)^{2} \nonumber \\&\hspace{5 em}-\frac{k_B T  w^2}{a_0^D r^2} \left(  P_{ij}^{T} \partial_{i} \*f \cdot \partial_{j} \*f \right)^2\bigg].
\end{align}
Thus, at this order of approximation, the coupling results in a shift in the coefficient of one of the terms already present in the effective Hamiltonian for a pristine membrane. Such a shift is not expected to affect flexural phonon critical exponents. To emphasize this point, the projection operator terms are rewritten in terms of mutually orthogonal tensors $M_{ij,kl}$ and $N_{ij,kl}$, defined in Appendix A \cite{le2018anomalous}.
\begin{align}
\mathcal{H} &\approx \int^\prime d^D x \bigg[\frac{\kappa}{2}|\nabla^2 \*f|^2\nonumber +\frac{\mu}{4} M_{ij,kl} \left( \partial_i \*f \cdot \partial_{j} \*f\right)\left(\partial_{k} \*f \cdot \partial_{l} \*f\right)\nonumber
\\&+ \frac{1}{4}\left(b-4(D-1) \frac{k_B T  w^2}{a_0^D r^2}\right) N_{ij,kl}\left( \partial_i \*f \cdot \partial_{j} \*f\right)\left(\partial_{k} \*f \cdot \partial_{l} \*f\right).
\end{align}
For $D=2$, $M_{ij,kl}=0$ and the Young's modulus $Y=2b$, allowing the influence of the coupling to magnetization to be written as an effective Young's modulus,
\begin{equation}
Y_{eff}=Y-\frac{8 k_B T  w^2}{a_0^D r^2}.
\end{equation}

\section{PERTURBATIVE RENORMALIZATION GROUP}
\label{sec:rg}
Next, an $\epsilon=4-D$ expansion is performed about the upper critical dimension for both the Ising model and the membrane model, $D_{\mathrm{uc}}=4$, using the energy functional given by Eq. \ref{eq:energy}. No new physical fixed points appear at one-loop order for $d_c\equiv d-D \leq 12$. At the uncoupled fixed point corresponding to the most stable fixed point of an isolated thermalized membrane and an isolated compressible Ising model, $w$ is a relevant operator for $d_c<12$. 


Following standard techniques \cite{kardar}, $m$ and $\*f$ are separated into long wavelength components with ${0 < |\*q| < \Lambda/e^{\delta l}}$ and short wavelength components with ${\Lambda/e^{\delta l} < |\*q| < \Lambda}$. Then, the Hamiltonian is coarse-grained by integrating over the short wavelength components, keeping terms at one-loop order. Upon rescaling $\*q$, renormalizing $m$ and $\*f$, and taking $\delta l \rightarrow 0$, the following recursion relations are derived. Further details are provided in Appendix B. Note that a factor of $\beta=1/k_B T$ has been absorbed into the coupling constants to simplify expressions.
\begin{align}
\frac{d r}{dl} &= 2 r +  \frac{(3 \tilde{u}+ v)\Lambda^4}{2\pi^2 (K  \Lambda^2 +r)},\label{eq:wrecursionr}\\
\frac{d \tilde{u}}{dl} &= \epsilon \tilde{u} - \frac{9 \tilde{u}^2 \Lambda^4}{2\pi^2(K\Lambda^2+r)^2} - \frac{5 d_c w^2 }{64 \pi^2 \kappa^{2}},\\
\frac{d v}{d l}&= \epsilon v -  \frac{3 \tilde{u} v \Lambda^4}{\pi^2(K \Lambda^2+r)^2}-\frac{v^2  \Lambda^4}{2\pi^2(K\Lambda^2+r)^2},\\
\frac{db}{dl}&=\epsilon b-\frac{5(4+d_c) b^2}{192 \pi^2\kappa^2}- \frac{5 b\mu}{24 \pi^2\kappa^2}-\frac{3w^2 \Lambda^4}{2 \pi^2(K \Lambda^2+r)^2},\\
\frac{d\mu}{dl}&=\epsilon \mu -\frac{5\mu b}{48\pi^2 \kappa^2}-\frac{(20+d_c)\mu^2}{96\pi^2 \kappa^2},\\
\frac{dw}{dl}&=\epsilon w-\frac{5(d_c+2) wb}{192 \pi^2\kappa^2}-\frac{5 w\mu}{48 \pi^2\kappa^2}-\frac{3 \tilde{u}w \Lambda^4}{2 \pi^2 (K \Lambda^2 + r)^2}.
\label{eq:wrecursionw}
\end{align}
These equations have 32 fixed points. Sixteen of the fixed points have $w^*=0$ and are thus uncoupled, emerging from the product of the four fixed points found by \citet{sak1974critical} and the four fixed points found by \citet{aronovitz} with $w=0$. The remaining sixteen coupled fixed points are unique to this model.

The uncoupled fixed points have a simple and familiar form, so they are listed explicitly and examined first. To lowest order in $\epsilon$, the four fixed points for $(r,\tilde{u},v)$ are $(0,0,0), (-\frac{K\Lambda^2 \epsilon}{2}, 0, 2K^2 \pi^2 \epsilon),$ $ (-\frac{K\Lambda^2 \epsilon}{3}, \frac{2 K^2 \pi^2 \epsilon}{9},  \frac{2 K^2 \pi^2 \epsilon}{3}), (-\frac{K\Lambda^2 \epsilon}{6}, \frac{2 K^2 \pi^2 \epsilon}{9},0)$. The four fixed points for $(b,\mu)$ are $(0,0), (0, \frac{96 \pi^2 \kappa^2 \epsilon}{20+d_c}),$ $(\frac{192 \pi^2 \kappa^2 \epsilon}{5(4+d_c)}, 0), (\frac{192 \pi^2 \kappa^2 \epsilon}{5(24+d_c)},\frac{96 \pi^2 \kappa^2 \epsilon}{24+d_c})$. 

First, the behavior of the system in the $b-\mu$ plane is considered for the uncoupled fixed points. When ${D<4}$, the membrane will be controlled by the Aronovitz-Lubensky fixed point (${b^*=\frac{192 \pi^2 \kappa^2 \epsilon}{5(24+d_c)}, \mu^*=\frac{96 \pi^2 \kappa^2 \epsilon}{24+d_c}}$). This fixed point has two irrelevant directions, allowing the system to tune itself to criticality at long wavelengths \cite{aronovitz}. The behavior of the system in the $\tilde{u}-v$ plane is more subtle. Only one of the fixed points has two irrelevant directions in this plane for $D<4$: $\tilde{u}^*= \frac{2 K^2 \pi^2 \epsilon}{9}, v^*=\frac{2 K^2 \pi^2 \epsilon}{3}$. However, this fixed point is inaccessible: Recall that $v$ must be negative since the bulk and shear modulus are required to be positive. Therefore, as discussed in \citet{sak1974critical}, a system starting with $v<0$ may flow towards the Wilson-Fisher fixed point $(\tilde{u}^*= \frac{2 K^2 \pi^2 \epsilon}{9},  v^*=0)$ depending on initial parameters since $\lambda_{\tilde{u}}=-\epsilon<0$ but will then be pushed towards a line of instability at $\tilde{u}+v=0$ by the relevance of $v$ ($\lambda_v=3\epsilon/9$). If $|v|$ is small initially, corresponding to a stiff lattice/weak coupling, the system may be controlled by the Wilson-Fisher fixed point for a wide range of temperatures and display the same critical behavior as a rigid Ising model there. Note that \citet{sak1974critical} also provides an exponent relation argument that suggests $\lambda_v$ is given by critical exponents $\alpha/\nu$ when higher order terms are included. For a $D=2$ system, Onsager's exact solution can be substituted, indicating that $v$ is a marginal perturbation with $\lambda_v=0$.

Thus, among the uncoupled fixed points, the product of the Aronovitz-Lubensky and Wilson-Fisher fixed points is the most stable. To assess the relevance of $w$ at this fixed point, recursion relations are linearized.
{\small
\begin{equation}
\frac{d \delta w}{d l} = \left( \epsilon- \frac{5 (d_c +2) b^*}{192 \pi^2 \kappa^2} - \frac{5 \mu^*}{48 \pi^2 \kappa^2} - \frac{3 \tilde{u}^* \Lambda^4}{2 \pi^2(K \Lambda^2 +r)^2}\right)\delta w, \nonumber
\end{equation}}
\vspace{-1 em}
\begin{equation}
\lambda_w=\left( \frac{12-d_c}{3(24+d_c)}\right) \epsilon.
\end{equation}
This calculation shows that at the Wilson-Fisher-Aronovitz-Lubensky fixed point, $w$ is a relevant operator for $d_c < 12$. For $d_c = 1,\epsilon=2$ the eigenvalue of $w$ is $22/75\approx 0.3$. 

The remaining sixteen fixed points that have nonzero $w$ are now considered. When $d_c<2$, all nonzero fixed point values of $w$ are purely imaginary. Four fixed points become real and nonzero when $d_c>2$. However, all of these fixed points have $\mu^*=0$ and are thus unphysical. Four additional fixed points become real and nonzero when $d_c>12$. Two of these fixed points are unphysical with $v^*>0$. The two remaining fixed points have $v=0$. The expressions for the eigenvalues of these fixed points are cumbersome so they are solved for numerically with $d_c$ between 13 and 1000. For all values tested, the two new physical fixed points have three positive eigenvalues, making them less stable than the Wilson-Fisher-Aronovitz-Lubensky fixed point, for which only $\lambda_r$ and $\lambda_v$ are positive when $d_c>12$. 

In Appendix C, recursion relations for a flexural $n$-component magnetic system are given. These equations reveal that increasing $n$ at fixed $d_c$ can also cause $w$ to become irrelevant and that the Wilson-Fisher-Aronovitz-Lubensky fixed point remains more stable than any coupled fixed points that arise.




\section{DISCUSSION}
\label{discussion}
A renormalization group analysis finds that the most stable physical fixed point of a flexural Ising model is uncoupled, formed by the product of a Wilson-Fisher Ising fixed point and an Aronovitz-Lubensky membrane fixed point. However, despite being more stable than alternatives, this fixed point nonetheless has three relevant directions for $d=3, D=2$. Significantly, the coupling between flexural phonons and magnetization is relevant when codimension $d-D< 12$. This relevant coupling could lead to new physics for the flexural Ising model at long wavelengths as flexural and magnetic degrees of freedom become more strongly coupled, a possibility hinted at by the positive value of the specific heat exponent $\alpha$ found in simulations of a closely related discrete model \cite{hanakata2022anomalous}. However, since the present calculation does not reveal a new nontrivial fixed point, it is unable to make predictions about what new critical behavior is expected due to this relevant coupling. It may be of interest to repeat this analysis with more sophisticated techniques \cite{le2018anomalous, coquand2020flat, doussal2023tattered} since it is possible that alternative methods could either reveal new fixed points or find that the flexural phonon-magnetization coupling is in fact an irrelevant operator when higher order terms are taken into account. An irrelevant coupling would be consistent with measurements of critical exponents $\gamma, \nu,$ and $\beta$ in the buckled dilation array model. It would also be interesting to study the effect of curvature on the flexural Ising model, which enters as an effective external field \cite{plummer2022curvature}. 

\begin{acknowledgments}
It is a pleasure to acknowledge many stimulating discussions with David R. Nelson and to thank him for his valuable suggestions. I am also grateful to Paul Hanakata for enlightening comments. In addition, I thank Suraj Shankar, Pierre Le Doussal, and Leo Radzihovsky for their insights. 
\end{acknowledgments}

\appendix
\section*{APPENDIX A: INTEGRATING OVER IN-PLANE PHONONS}
To integrate out in-plane phonons, the strain tensor is separated into $\*q=0$ and $\*q \neq 0$ components with the conventions $\xi (\*x)= \sum_{\*q}\xi(\*q) e^{i\*q \cdot \*x}$ and ${\xi (\*q)= \frac{1}{L^D} \int d^D x \xi(\*x) e^{-i\*q \cdot \*x}}$, where $L^D$ is the $D$-dimensional membrane area. 
\begin{equation}
u_{i j}(\mathbf{x})=u_{i j}^{0}+A_{i j}^{0}+\sum_{\mathbf{q} \neq 0}\left(\frac{i}{2} q_{i} u_{j}(\mathbf{q})+\frac{i}{2} q_{j} u_{i}(\mathbf{q})+A_{i j}(\mathbf{q})\right) e^{i \mathbf{q} \cdot \mathbf{x}},
\end{equation}
with $A_{ij}(\*x)= \frac{1}{2}\frac{\partial \*f}{\partial x_i} \cdot \frac{\partial \*f}{\partial x_j}$, the nonlinear part of the strain tensor. The terms $A_{ij}^0$ and $u_{ij}^0$ are the uniform strains that do and do not depend on $\*f$, respectively. 
The terms in Eq. \ref{eq:totalen} with contributions from the $\*q=0$ mode are
\begin{align}
\mathcal{H}_0&= \mu L^D (u_{ij}^0+A_{ij}^0)^2\nonumber \\ &+\frac{\lambda L^D}{2} (u_{kk}^0+A_{kk}^0)^2+g (u_{kk}^0+A_{kk}^0) \int d^D x m(\*x)^2 ,
\end{align}
Defining $w_{ij}^0=u_{ij}^0+A_{ij}^0+ \frac{g}{L^D(2\mu+D\lambda)}\delta_{ij}\int d^D x m(\*x)^2$ completes the square and allows for integration over $w_{ij}^0$. Following this procedure, one term remains:
\begin{align}
\mathcal{H}_0&= -\frac{D g^2}{2 L^D (2 \mu +D \lambda)}\left(\int d^D x m(\*x)^2 \right)^2\nonumber \\&= -\frac{g^2}{2 L^D B}\left(\int d^D x m(\*x)^2 \right)^2,
\end{align}
where $B$ is the $D$-dimensional bulk modulus. 

Next, consider $\*q \neq 0$. The terms in Eq. \ref{eq:totalen} with contributions from the $\*q\neq 0$ modes are
{\small
\begin{align}
\frac{\mathcal{H}_{\neq 0}}{L^D}&=\sum_{\*q \neq 0}\Big[ \frac{1}{2}u_i(-\*q) \left[\mu q^2 P_{ij}^T(\*q)+ (2\mu+\lambda) q^2P_{ij}^L(\*q) \right] u_j(\*q)\nonumber \\
 &\hspace{2 em}-u_i(-\*q) \left[2i\mu q_j A_{ij}(\*q)+i\lambda q_i A_{kk}(\*q)+ i g q_i \Psi(\*q)\right] \nonumber \\
&\hspace{2 em}+gA_{kk}(\*q) \Psi(-\*q)+ \mu |A_{ij}(\*q)|^2 + \frac{\lambda}{2} |A_{kk}(\*q)|^2\Big],
\end{align}}
with 
\begin{align*}
P_{ij}^L(\*q) &= \frac{q_i q_j}{q^2},  \\
P_{ij}^T(\*q) &= \delta_{ij} - \frac{q_i q_j}{q^2},  \\
\Psi(\*q)&= \frac{1}{L^D} \int d^Dx m(\*x)^2 e^{-i \*q \cdot \*x}.
\end{align*}

As above, terms are integrated by defining ${w_i = u_i - C_{ik}^{-1}j_k}$ with
\begin{align}
C_{ij}^{-1}(\*q)&= \frac{1}{\mu q^2}P_{ij}^T(\*q) + \frac{1}{(2\mu +\lambda) q^2} P_{ij}^L (\*q),\nonumber \\
j_i(\*q)&=2i \mu q_j A_{ij}(\*q)+ i \lambda q_i A_{kk}(\*q) + i g q_i \Psi(\*q). \nonumber 
\end{align}
After integrating over $w_i(\*q)$, the remaining terms are
{\small
\begin{align}
&\frac{\mathcal{H}_{\neq 0}}{L^D}= \sum_{\*q \neq 0} \Big[-\frac{2 \mu}{q^2} q_i  A_{ij}(-\*q) q_kA_{kl}(\*q)\nonumber \\ &\hspace{2 em}+\frac{2 \mu(\lambda+\mu)}{(2 \mu+\lambda)q^4}q_i q_k A_{ik}(-\*q)q_j q_l A_{jl}(\*q)+ \mu |A_{ij}(\*q)|^2 \nonumber \\ &\hspace{2 em}-\frac{ 2\lambda \mu }{(2\mu+\lambda) q^2}q_j q_i A_{ij}(-\*q)A_{kk}(\*q)+ \frac{\lambda\mu}{2\mu+\lambda} |A_{kk}(\*q)|^2
\nonumber\\&\hspace{2 em}-\frac{2 g \mu}{(2\mu+\lambda)q^2} q_i q_j A_{ij}(\*q)\Psi(-\*q)\nonumber\\
&\hspace{2 em}+\frac{2 g \mu}{(2\mu+\lambda)}A_{kk}(\*q) \Psi (-\*q)-\frac{g^2}{2(2\mu+\lambda)}|\Psi(\*q)|^2\Big].
\end{align}
}

Following \citet{le2018anomalous}, this expression can be simplified using projection operators, expressing $A_{ij}(\*q)=\-\frac{1}{2}\sum_{\*q_1,\*q_2} q_{1 i} q_{2 j} \*f(\*q_1) \cdot \*f(\*q_2) \delta_{\*q_1+\*q_2, \*q}$, and combining all terms quartic in $\*f$ by defining a fourth-order tensor $R_{ij, kl}(\*q)$. 
{\small
\begin{align}
&\frac{\mathcal{H}_{\neq 0}}{L^D}=\sum_{\*q \neq 0} \Big[\mu\left(P_{i l}^{T}(\*q) A_{i j}(\*q)\right)\left(P_{j k}^{T}(\*q) A_{k l}(-\*q)\right)\nonumber \\&+\frac{\mu \lambda}{2 \mu+\lambda}\left(P_{ij}^{T}(\*q) A_{ij}(\*q)\right)\left(P_{kl}^{T}(\*q) A_{kl}(-\*q)\right) \nonumber \\
&+ \frac{2 g \mu}{2 \mu+\lambda} P_{ij}^T(\*q) A_{ij}(\*q) \Psi(-\*q) - \frac{g^2}{2(2 \mu+\lambda)} |\Psi(\*q)|^2\Big], \nonumber \\
=&\frac{1}{4}\sum\limits_{\substack{ \*q_1+\*q_2 =\*q \neq 0 \\ \*q_3+\*q_4 =-\*q \neq 0}}\hspace{-1.5 em}R_{ij, kl}(\*q) q_{1 i} q_{2 j} q_{3 k} q_{4 l} \left(\*f(\*q_1) \cdot \*f(\*q_2) \right)\left(\*f(\*q_3) \cdot \*f(\*q_4) \right) \nonumber \\
 &-\frac{g \mu}{2 \mu+\lambda}\sum_{\*q_1+\*q_2=\*q \neq 0} P_{ij}^T(\*q) q_{1 i} q_{2 j} \left(\*f(\*q_1) \cdot \*f(\*q_2) \right)\Psi(-\*q)\nonumber \\& -\frac{g^2}{2(2 \mu+\lambda)} \sum_{\*q \neq 0} |\Psi(\*q)|^2, 
\end{align}
}
with
{\small
\begin{align}
&R_{ij, kl}=\mu M_{ij, kl}(\*q)+ b N_{ij, kl}(\*q), \nonumber\\
&=\mu\left(\frac{1}{2} \left( P_{i l}^T(\*q) P_{j k}^T(\*q)+   P_{i k}^T(\*q) P_{j l}^T(\*q)\right)- \frac{1}{D-1}P_{i j}^T(\*q)P_{k l}^T(\*q) \right) \nonumber \\&\hspace{3 em}+ \frac{b}{D-1}P_{i j}^T(\*q)P_{k l}^T(\*q), \nonumber
\end{align}
}where $b= \frac{\mu (2\mu+D \lambda)}{2 \mu+\lambda}$ is a coupling constant proportional to the $D$-dimensional bulk modulus. In $D=2$, $2b$ is equivalent to the $2D$ Young's modulus and $M_{ij,kl}=0$ \cite{le1992self, le2018anomalous}.

Next, terms that depend on the $\*q=0$ mode of the strain tensor and terms independent of the strain tensor are added back in. In Fourier space, the total energy is
{\small
\begin{align}
&\frac{\mathcal{H}}{L^D}= \sum_{\mathbf{q}}\Big[\frac{\kappa}{2} q^{4} |\*f(\mathbf{q})|^{2}+\left(\frac{r+Kq^2}{2}\right)|m(\*q)|^2 \Big]\nonumber \\&-\frac{g \mu}{2 \mu+\lambda} \sum\limits_{\substack{\*q_1+\*q_2=\*q \neq 0 \\ \*q_3+\*q_4 = -\*q }} \hspace{-1 em} P_{i j}^T(\*q_1 + \*q_2) q_{1 i} q_{2 j}(\*f(\*q_1) \cdot \*f(\*q_2))m(\*q_3) m(\*q_4)\nonumber \\
&+\frac{1}{4} \sum\limits_{\substack{ \*q_1+\*q_2 =\*q \neq 0 \\ \*q_3+\*q_4 =-\*q}} \hspace{-1 em} R_{ij, kl}(\*q)q_{1 i} q_{2 j} q_{3 k} q_{4 l}  (\*f(\*q_1) \cdot \*f(\*q_2)) ( \*f(\*q_3)\cdot \*f(\*q_4))\nonumber \\
&+ \left(u-\frac{g^2}{2(2\mu+\lambda)}\right) \sum\limits_{\*q_1+\*q_2+\*q_3+\*q_4=0}\hspace{-1 em} m(\*q_1)m(\*q_2)m(\*q_3)m(\*q_4)\nonumber \\
&+\frac{g^2}{2}\left( \frac{1}{2\mu+\lambda} -\frac{D}{2\mu+D\lambda} \right) \left(\sum_{\*q}|m(\*q)|^2\right)^2.
\label{eq:ftenergy}
\end{align}
}
Note that terms quartic in $m$ have been rearranged so that there is one restricted and one unrestricted sum over wavevectors, following \citet{bruno1980renormalization}.

In real space, this expression can be written
{\small
\begin{align}
\mathcal{H}&= \frac{g^2}{2L^D}\left( \frac{1}{2 \mu+\lambda}-\frac{D}{ 2\mu+D\lambda}\right) \left(\int d^D x m^2 \right)^2\nonumber \\&+ \int d^D x \bigg[ \frac{\kappa}{2} |\nabla^2 \*f |^2+ \frac{K}{2}\left(\nabla m\right)^{2}+\frac{r}{2} m^{2}+\left(u- \frac{g^2}{2(2\mu+\lambda)}\right) m^{4} \bigg]\nonumber \\
&+\int^\prime d^D x \bigg[\frac{\mu}{4}\left(P_{il}^{T} \partial_{i} \*f \cdot \partial_{j} \*f\right)\left(P_{jk}^{T} \partial_{k} \*f \cdot \partial_{l} \*f\right)\nonumber \\&\hspace{1 em}+\frac{\mu \lambda}{4(2 \mu+\lambda)}\left(P_{ij}^{T} \partial_{i} \*f \cdot \partial_{j} \*f\right)^{2} +  \frac{g \mu}{2 \mu+\lambda} \left( m^2 P_{ij}^T \partial_i \*f\cdot \partial_j \*f \right) \bigg],
\end{align}
}where primed integrals signify that the zero mode is excluded. 

Specializing to $D=2, d=3$, the case relevant to Ising-like buckled dilation arrays and atomically thin materials, Eq. \ref{eq:energy} simplifies to \cite{hanakata2022anomalous}
{\small
\begin{align}
&\mathcal{H}= \frac{g^2}{2L^2}\left( \frac{1}{2 \mu+\lambda}-\frac{1}{ \mu+\lambda}\right) \left(\int d^2 x m^2 \right)^2\nonumber\\ 
&+ \int d^2 x \bigg[ \frac{\kappa}{2} |\nabla^2 f|^2 + \frac{K}{2}\left(\nabla m\right)^{2}+\frac{r}{2} m^{2}+\left(u- \frac{g^2}{2(2\mu+\lambda)}\right) m^{4} \bigg]\nonumber \\
&+\int^\prime d^2 x \left[ \frac{Y}{8} \left(P_{ij}^T \partial_i f \partial_j f \right)^2 +  \frac{g \mu}{2 \mu+\lambda} \left( m^2 P_{ij}^T \partial_i f \partial_j f \right) \right] .
\end{align}
}
\section*{APPENDIX B: $\epsilon$-EXPANSION CALCULATION}
An $\epsilon$-expansion has already been performed for both the compressible Ising model \cite{sak1974critical, bruno1980renormalization} and for the crystalline membrane \cite{aronovitz}. 
Therefore, this appendix focuses on the coupling between flexural phonons and magnetization proportional to $w$ in Eq. \ref{eq:energy}. When $w=0$, the system reduces to an uncoupled compressible Ising model and crystalline membrane and our results are consistent with past work, as expected. 

Four new diagrams contribute to the renormalization group recursion relations when $w$ is included: 
\begin{enumerate}
\item A correction to $w$ from the product of the interactions proportional to $w$ and $\tilde{u}$.
\item A correction to $w$ from the product of the interactions proportional to $w$ and $b$.
\item A correction to $\tilde{u}$ from the product of two interactions proportional to $w$.
\item A correction to $b$ from the product of two interactions proportional to $w$.
\end{enumerate}

Calculations of the first two corrections are given below; other calculations are similar. 

The correction to $w$ from the $w$ and $\tilde{u}$ interaction product has a multiplicity of 12 and enters with a minus sign, as it is second order in the expansion around the Gaussian model. Written in Fourier space as in Eq. \ref{eq:ftenergy}, this term is
{\small
\begin{align*}
&12\tilde{u}wL^{2D} \sum\limits_{\substack{ \*q_1...\*q_4\\ \*p_1...\*p_4}}P^T_{i j}(\*q)q_{1 i} q_{2 j} \left(\*f^<(\*q_1)\cdot f^<(\*q_2)\right)  m^<(\*p_3)m^<(\*p_4)\\
&\times \langle m^>(\*q_3)m^>(\*p_1)\rangle \langle m^>(\*q_4)m^>(\*p_2) \rangle \delta_{\sum_i\*q_i,0}\delta_{\sum_i\*p_i,0}
\end{align*}
}
Superscripts $<$ and $>$ denote long and short wavelength modes respectively. Averages are taken over short wavelength modes using the propagator ${\langle m^>(\*q)m^>(\*q^\prime) \rangle=\frac{\delta_{\*q, -\*q^\prime}}{L^D ( K q^2 +r) }}$. 
{\small
\begin{align*}
&12\tilde{u}wL^{D} \hspace{-1 em} \sum_{\*q_1,\*q_2,\*p_3,\*p_4}\hspace{-1 em} P^T_{i j}(\*q)q_{1 i} q_{2 j}  \left(\*f^<(\*q_1)\cdot f^<(\*q_2)\right) m^<(\*p_3)m^<(\*p_4)\\
&\times \frac{\delta_{\*q_1+\*q_2+\*p_3+\*p_4,0}}{L^D}\hspace{-1 em}\sum_{\Lambda/e^{\delta l}<\*k< \Lambda} \frac{1}{(K k^2+r)(K (\*k+\*q_1+\*q_2)^2+r)}
\end{align*}
}
By assuming $|\*q_1|$ and $|\*q_2|$, both long wavelength modes, are much smaller than $|\*k|$, a short wavelength mode, the sum can be expanded. Upon neglecting higher-order terms and integrating $k$ over a thin shell of momentum space, this term becomes
{\small
\begin{align}
12\tilde{u}wL^{D}& \hspace{-1 em}\sum_{\*q_1,\*q_2,\*p_3,\*p_4}\hspace{-1 em}P^T_{i j}(\*q)q_{1 i} q_{2 j}  \left(\*f^<(\*q_1)\cdot f^<(\*q_2)\right) m^<(\*p_3)m^<(\*p_4)\nonumber\\ 
&\times  \delta_{\*q_1+\*q_2+\*p_3+\*p_4,0}\frac{S_{D} \Lambda^D}{(2\pi)^D(K \Lambda^2 + r)^2} (\delta l),\nonumber
\end{align}
}where $S_D$ is the surface area of a $D$-dimensional unit sphere.

The correction to $w$ from the $w$ and $b$ interaction product has a multiplicity of 4 and also enters with a minus sign. This interaction is proportional to the tensor $N_{ij,kl}$. The term proportional to $\mu M_{ij,kl}$ does not provide a correction. 
\begin{align*}
&\frac{w b L^{2D}}{D-1}\sum\limits_{\substack{ \*q_1...\*q_4  \\ \*p_1...\*p_4}} P_{kl}^T(\*p_1+\*p_2)p_{1 k} p_{2 l}\left(\*f^<(\*p_1)\cdot f^<(\*p_2)\right)\\&\times  P_{mn}^T(\*p_3+\*p_4)  p_{3 m} p_{4 n} m^<(\*q_3)m^<(\*q_4)P^T_{i j}(\*q_1+\*q_2)q_{1 i} q_{2 j}\\& \times   
\langle f_\alpha ^>(\*q_1) f_\beta ^>(\*p_3) \rangle \langle f_\alpha ^>(\*q_2) f_\beta ^>(\*p_4) \rangle \delta_{\sum_i \*q_i,0} \delta_{\sum_i \*p_i,0}
\end{align*}
The propagator for flexural phonon modes is $\langle f^>_\alpha (\*q) f^>_\beta (\*q^\prime)\rangle =\frac{\delta_{\*q, -\*q^\prime} \delta_{\alpha \beta}}{L^D \kappa q^4}$. 
{\small
\begin{align*}
\frac{w b d_c L^{D}}{D-1}\sum\limits_{\substack{ \*q_3,\*q_4  \\ \*p_1, \*p_2}} & m^<(\*q_3)m^<(\*q_4)P_{kl}^T(\*p_1+\*p_2)p_{1 k} p_{2 l}\left(\*f^<(\*p_1)\cdot f^<(\*p_2)\right)\\
 &\times \delta_{\*q_3+\*q_4+\*p_1+\*p_2,0}\sum_{\*q_1, \*q_2}  
\frac{\left(P^T_{i j}(\*q_1+\*q_2)q_{1 i} q_{2 j}\right)^2}{ L^D \kappa^2 q_1^4 q_2^4} \nonumber
\end{align*}}
The sum over short wavelength modes can be integrated after simplifying with the substitution $\*k=\*q_1$, $\*q_2=\*q-\*k$.
\begin{align*}
&\frac{1}{(2\pi)^D}\int_{\Lambda/e^{\delta l}}^{\Lambda} d^D k
\frac{\left(P^T_{i j}(\*q)k_{i} k_{j}\right)^2}{\kappa^2 k^4 (\*q-\*k)^4}\\&=\frac{1}{(2\pi)^D}\int_{\Lambda/e^{\delta l}}^{\Lambda} d^D k
\frac{\left(1- (\hat{\*k} \cdot \hat{\*q})^2 \right)^2}{\kappa^2 (\*q-\*k)^4}\\
&\approx\frac{1}{(2\pi)^D}\int_{\Lambda/e^{\delta l}}^{\Lambda} d^D k
\frac{\left(1- (\hat{\*k} \cdot \hat{\*q})^2 \right)^2}{\kappa^2 k^4} \approx \frac{S_D \Lambda^{D-4}(D^2-1)}{(2\pi)^D \kappa^2 D(D+2)} \delta l.
\end{align*}

In each of these terms, wavevectors are rescaled by $e^{\delta l}$ and magnetization and flexural phonon fields are renormalized by $e^{\zeta_I \delta l}$ and $e^{\zeta_f \delta l}$ respectively.
\begin{align}
w^\prime=e^{\delta l(D+2 \zeta_I + 2 \zeta_f-2)} \bigg(&w - 12 \tilde{u}w \frac{S_D \Lambda^D}{(2\pi)^D (K \Lambda^2+r)^2} \delta l \nonumber\\
&-  b w \frac{d_c(D+1)}{D(D+2)}\frac{S_{D} \Lambda^{D-4}}{(2 \pi)^{D}\kappa^{2}} \delta l\bigg).
\end{align}
Differential recursion relations are derived upon expanding $e^{\delta l} \approx 1+ \delta l$. The remaining recursion relations are derived by following the same steps for the other interactions.
\begin{widetext}
\begin{align}
\frac{d r}{dl} &= (D+2\zeta_I) r + (12 \tilde{u}+4 v) \frac{\Lambda^D S_{D}}{(2\pi)^D(K  \Lambda^2 +r)},\\
\frac{dK}{dl}&=(D+2 \zeta_I-2)K,\\
\frac{d \tilde{u}}{dl} &= (D+4 \zeta_I) \tilde{u} - 36 \tilde{u}^2 \frac{S_{D} \Lambda^D}{(2\pi)^D(K\Lambda^2+r)^2} - w^2\frac{d_c (D^2-1)}{D(D+2)} \frac{S_{D} \Lambda^{D-4}}{(2 \pi)^{D}\kappa^{2}},\\
\frac{d v}{d l}&= (D+4 \zeta_I) v - 24 \tilde{u} v \frac{S_{D} \Lambda^D}{(2\pi)^D(K \Lambda^2+r)^2}-4 v^2 \frac{S_{D} \Lambda^D}{(2\pi)^D(K\Lambda^2+r)^2},\\
\frac{d \kappa}{d l } &=(D+2\zeta_f-4)\kappa + (2b+2\mu(D-2))\frac{D+1}{D(D+2)}\frac{S_{D} \Lambda^{D-4}}{(2 \pi)^{D}\kappa}, \\
\frac{d b}{d l } &=  (D+4\zeta_f-4) b- b^{2}\frac{ d_{c}(D+1)}{D(D+2)}\frac{ S_{D} \Lambda^{D-4}}{(2 \pi)^{D}\kappa^{2}}- 4 w^2\frac{(D-1) S_{D}\Lambda^D}{(2\pi)^D (K \Lambda^2+r)^2},\\
\frac{d \mu}{d l } &=(D+4\zeta_f-4) \mu- 2\mu^{2}\frac{d_{c}}{D(D+2)}\frac{ S_{D} \Lambda^{D-4}}{(2 \pi)^{D}\kappa^{2}}, \\
\frac{d w}{d l } &= (D+2\zeta_f+ 2\zeta_I-2)w -12\tilde{u}w \frac{S_{D} \Lambda^D}{(2\pi)^D(K \Lambda^2 + r)^2}-b w \frac{d_c(D+1)}{D(D+2)}\frac{S_{D} \Lambda^{D-4}}{(2 \pi)^{D}\kappa^{2} }.
\end{align}
\end{widetext}
\vspace{-1 em}
Parameters $\zeta_I$ and $\zeta_f$ are chosen to fix $K$ and $\kappa$.
\begin{align*}
\zeta_I&= 1-\frac{D}{2},\nonumber \\
\zeta_f&= \frac{4-D}{2} -(b+\mu(D-2))\frac{D+1}{D(D+2)}\frac{S_{D} \Lambda^{D-4}}{(2 \pi)^{D}\kappa^2}.
\end{align*}
The expressions for $\zeta_I$ and $\zeta_f$ are used to rewrite the recursion relations with $S_{D}/(2\pi)^D \approx \frac{1}{8 \pi^2}$ near $D=4$, neglecting higher order terms in $\epsilon=4-D$. Following these simplifications, Eqs. \ref{eq:wrecursionr}-\ref{eq:wrecursionw} are derived.

\vspace{-1.5 em}
\section*{APPENDIX C: $n$-COMPONENT MAGNET}
If an $n$-dimensional magnetization is considered instead, the recursion relations become
\begin{align}
\frac{d r}{dl} &= 2 r +  \frac{((n+2) \tilde{u}+ n v)\Lambda^4}{2\pi^2 (K  \Lambda^2 +r)},\\
\frac{d \tilde{u}}{dl} &= \epsilon \tilde{u} - \frac{(n+8) \tilde{u}^2 \Lambda^4}{2\pi^2(K\Lambda^2+r)^2} - \frac{5 d_c w^2 }{64 \pi^2 \kappa^{2}},\\
\frac{d v}{d l}&= \epsilon v -  \frac{(n+2) \tilde{u} v \Lambda^4}{\pi^2(K \Lambda^2+r)^2}-\frac{n v^2  \Lambda^4}{2\pi^2(K\Lambda^2+r)^2},\\
\frac{db}{dl}&=\epsilon b-\frac{5(4+d_c) b^2}{192 \pi^2\kappa^2}- \frac{5 b\mu}{24 \pi^2\kappa^2}-\frac{3 n w^2 \Lambda^4}{2 \pi^2(K \Lambda^2+r)^2},\\
\frac{d\mu}{dl}&=\epsilon \mu -\frac{5\mu b}{48\pi^2 \kappa^2}-\frac{(20+d_c)\mu^2}{96\pi^2 \kappa^2},\\
\frac{dw}{dl}&=\epsilon w-\frac{5(d_c+2) wb}{192 \pi^2\kappa^2}-\frac{5 w\mu}{48 \pi^2\kappa^2}-\frac{(n+2) \tilde{u}w \Lambda^4}{2 \pi^2 (K \Lambda^2 + r)^2}.
\label{eq:nrecursionw}
\end{align}
As before, sixteen fixed points are coupled and sixteen are uncoupled.
The uncoupled fixed points $(r,\tilde{u},v)$ for the magnetic system are $(0,0,0), (-\frac{K\Lambda^2 \epsilon}{2}, 0,\frac{ 2K^2 \pi^2 \epsilon}{n}),$ $ (-\frac{3 K\Lambda^2 \epsilon}{(n+8)}, \frac{2 K^2 \pi^2 \epsilon}{n+8},  \frac{2(4-n) K^2 \pi^2 \epsilon}{n(8+n)}),$ $ (-\frac{(n+2)K\Lambda^2 \epsilon}{2(8+n)}, \frac{2 K^2 \pi^2 \epsilon}{n+8},0)$, as found previously \cite{sak1974critical}. The eigenvalue of $w$ at the Wilson-Fisher-Aronovitz-Lubensky fixed point is
\begin{equation}
\lambda_w= \left(- \frac{n+2}{n+8} + \frac{12}{24+d_c} \right) \epsilon.
\end{equation}
The coupling $w$ is relevant at this fixed point when $D<4$ and $d_c< \frac{12 (4-n)}{n+2}$, or equivalently, $n< \frac{2(24-d_c)}{d_c+12}$. 

Similarly, for the sixteen coupled fixed points, behavior can be affected by increasing $n$, $d_c$, or $D$. For small $n$ and $d_c$, $w^*$ is purely imaginary. As before, there are two thresholds, now a function of both $n$ and $d_c$, after which some values of $w^*$ become real and nonzero. First, when $n> \frac{8-2 d_c}{2+d_c}$, four fixed points become real and nonzero. For $n=1$, this condition corresponds to the $d_c>2$ threshold discussed in Sec. \ref{sec:rg}. As before, these fixed points are unphysical, as they have $\mu^*=0$. Four more fixed points become real and nonzero when $\frac{2(24- d_c)}{12+d_c} < n < \frac{20+3 d_c}{5}$. For $n=1$, this is equivalent to the $d_c>12$ threshold discussed prior. As before, the coupled fixed points that satisfy $v\leq 0$ are less stable than the uncoupled Wilson-Fisher-Aronovitz-Lubensky fixed points for any tested values of $(n, d_c, \epsilon)$. Note that at the uncoupled fixed point $v$ becomes an irrelevant operator for $n>4$ \cite{sak1974critical}. 
\bibliographystyle{unsrtnat}
\bibliography{biblio}

\end{document}